**Toward a General Theory of Societal Collapse. A Biophysical Examination of Tainter's Model of the Diminishing Returns of Complexity.**

Ugo Bardi[1], Sara Falsini[2], and Ilaria Perissi[2]
1. Dipartimento di Chimica – Università di Firenze, Italy
and
2. Consorzio Interuniversitario per la Scienza e Tecnologia dei Materiali (INSTM)
Polo Scientifico di Sesto Fiorentino, via della Lastruccia 3, Sesto F. 50019 (Fi), Italy

**Abstract.** The collapse of large social systems, often referred to as "civilizations" or "empires," is a well-known historical phenomenon, but its origins are the object of an unresolved debate. In this paper, we present a simple biophysical model which we link to the concept that societies collapse because of the "diminishing returns of complexity" proposed by Joseph Tainter[1]. Our model is based on the description of a socio-economic system as a trophic chain of energy stocks which dissipate the energy potential of the available resources. The model produces various trajectories of decline, in some cases rapid enough that they can be defined as "collapses." At the same time, we observe that the exploitation of the resource stock ("production") has a strongly nonlinear relationship with the complexity of the system, assumed to be proportional to the size of the stock termed "bureaucracy." These results provide support for Tainter's hypothesis.[1]

**1. Introduction**

The collapse of large social systems, also called "civilizations" or "empires," is a well-known and highly studied subject. In many cases, the historical record does not provide quantitative data on these events, but in some cases it is possible to quantify the collapse phenomenon in terms, for instance, of the extent of the areas controlled by the central government as reported by Tageepera[2] or of the output of the economic system as reported by Sverdrup[3] and McCollen et al.[4] In these studies, we can observe how collapses are often rapid in comparison to the build-up of the social and economic structures of a civilization. This behaviour is consistent with Diamond's definition of collapse as, "*a drastic decrease in human population size and or political/economic/social complexity, over a considerable area, for an extended time*."[5]

Nevertheless, despite the number of studies in this area, there is little agreement on the causes of societal collapses and, in particular, on the possibility of a common mechanism causing them. Edward Gibbon was probably the first to attempt an interpretation of the fall of a large empire, the Roman one, attributing it mainly to the decline of the traditional values[6]. Later authors explained the fall of Rome in widely different ways and Demandt[7] (1984) lists about 210 different theories on this subject, probably an incomplete list. The same variety of interpretations affects the studies of the collapse of other societies in history, as described, for instance, by Tainter[8] in 2008.

No consensus appears to exist in this field but, overall, we can divide the interpretations of collapses into two main subsets: theories based on several independent causes (concauses) and theories based on a single cause that generates a cascade of different effects. An example of the first approach – several independent concauses – is the study by E. H. Cline on the collapse of the Late Bronze Age Mediterranean Civilization[9]. According to Cline, multiple negative effects occurred at the same time, including climate change, earthquakes, foreign invasions, and more. An extreme example of the multiplication of causes is the study Bury published in 1923[10] who argued that the collapse of the Roman Empire resulted from several contingent events all occurring at about the same time. Tainter comments[11] stating that Bury considers that "*The collapse was just bad luck*".

There are several examples of the second approach, single cause followed by a cascade of related events. One is Douglas Reynolds' interpretation of the fall of the Soviet Union in 1991[12]. Reynolds attributes it to mineral depletion and, specifically, to the cascade of negative effects generated by the growing costs of oil production which affected the whole Soviet economic system. Another single-factor model of civilization collapse factor has been proposed by Joseph Tainter in his study "The Collapse of Complex Societies"[1] and in later papers[8,13],[14]. Tainter identifies "diminishing returns," a well-known concept in economics, as the general factor in the decline and fall of civilizations. The idea is that, as societies become larger, more complex control structures are needed to maintain the cohesion of society and solve the problems that appear along their path. These structures can be described in terms of governments, the nobility, armies, bureaucracy, and the like. According to Tainter, as these structures become larger, they become less efficient, to the point that the economic returns they provide are smaller than their cost. At this point, society becomes unable to cope with the challenges it faces and must decline, or even collapse.

The contrast between single/multiple causes in the interpretation of the fall of societies highlights a general methodological problem. Not only data are often scarce on these historical phenomena, but their interpretation is often based on the author's personal judgment of the relative importance of the events he studies. It goes without saying that the collapse of civilizations is not amenable to experimental studies but, even taking this point into account, one may ask how proposing a specific interpretation of the fall of – say – the Roman Empire can be justified. Here, we have several problems, including the fact that the very concept of "causation" is hard to approach in a quantitative manner[15]. Nevertheless, we can choose to rely on the basic scientific concept that the preferable interpretation of an event is not only one that's compatible with the available data, but also which is of general validity – that is, can explain more than a single event of the same class. In this sense, Tainter's interpretation of "diminishing returns of complexity" provides a general framework to interpret a large number of cases and it is, therefore, an interesting idea in view of understanding the general phenomenon of societal collapse.

In the present study, we looked at Tainter's ideas using the modern concept of "System Science."[16] By using the modelling method known as "system dynamics"[17] we developed a simple biophysical model describing the evolution of a society. The model includes the effects of overshoot[18] and of diminishing returns in the exploitation of natural resources. It is not supposed to describe specific social systems but to provide a "mind-sized"[19] model the main factors that cause collapse. We find that the complexity of the system assumed to be proportional to the size of a stock such as "bureaucracy" follows a trajectory that makes the model compatible with the one proposed by Tainter. That is, the system shows a hysteresis that makes its trajectory non-reversible: reducing the costs of bureaucracy doesn't return society to the previous conditions of prosperity.

**2. Tainter's model:**

Tainter describes his model in his 1988 book "*The Collapse of Complex Societies*."[1] Here is an excerpt from the book.

> More complex societies are more costly to maintain than simpler ones, requiring greater support levels per capita. <..> It is the thesis of this chapter that return on investment in complexity varies, and that this variation follows a characteristic curve. More specifically, it is proposed that, in many crucial spheres, continued investment in sociopolitical complexity reaches a point where the benefits for such investment begin to decline, at first gradually, then with accelerated force. Thus, not only must a population allocate greater and greater amounts of resources to maintaining an evolving society, but after a certain point, higher amounts of this investment will yield smaller increments of return. Diminishing returns, it will be shown, are a recurrent aspect of sociopolitical evolution, and of investment in complexity.

The graphic representation of Tainter's mechanism is shown in figure 1 (redrawn from Tainter's book).

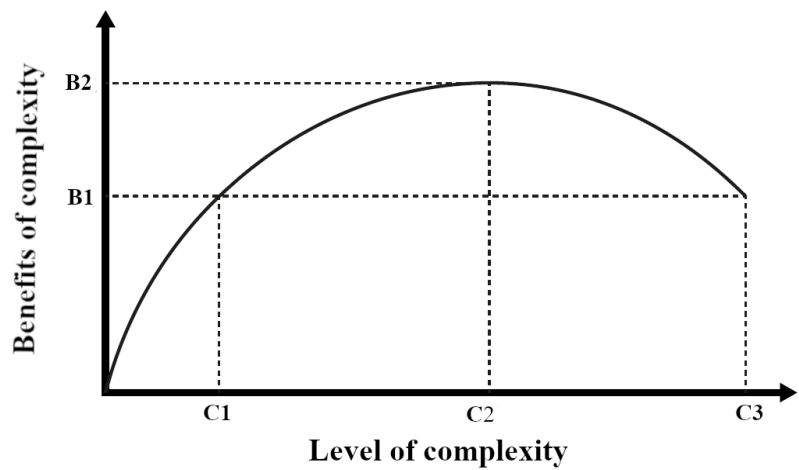

Figure 1: The diminishing returns to complexity, graphically described by Tainter[1]. (redrawn from Tainter's book).

Tainter's thesis is not directly based on quantitative data or models, but historical data are used to support it. For example, the decline of the content of silver in the Roman *denarius* for a period that goes from the 1st to the 3rd century AD is interpreted by Tainter as an indication that the Roman government was experiencing increasing financial difficulties. Tainter attributes this phenomenon to the increasing cost of the Roman bureaucracy, including the imperial court, resulting from the expansion of the Roman civilization during the 1st century BC and the 1st century AD.

Surprisingly, Tainter never mentions in his 1988 book the depletion of the Roman silver and gold mines in Spain, another possible cause of diminishing returns. The subject of mineral depletion in the ancient world is difficult to study for the lack of specific data on mineral production, but it is known that depletion was a problem for the Roman mines and that mining required progressively more efforts, for instance in terms of deeper mines[20]. Recent data show that the Roman mining of various metals and the Roman industrial activity collapsed together with the third century crisis.[4] These data don't necessarily imply that depletion was the problem, but do indicate that the lack of precious metals was due to a real decline in availability, not just to the expansion of the system. In a more recent discussion,[13] Tainter himself identifies the diminishing returns of mineral exploitation as related to the concept of "EROI" (energy returned for energy invested) developed by Hall and others[21]. The concept of EROI cannot be directly applied to the extraction of metals and other non-energy producing minerals, but the basic mechanism that leads to diminishing EROI for fossil fuels is valid for all mineral resources. The less costly resources are exploited first, and this leads to a gradual and irreversible increase in the energetic and monetary costs of extraction. This is one of the causes of diminishing returns in a complex society[22].

**3. Collapse: the systems science approach**

By developing the concept of "diminishing returns," Tainter links historical collapses to the modern concept of "systems science"[16] a field dedicated to the study of complex systems, all different, but all tending to show a similar behaviour in similar circumstances. Complex systems are often described in the framework of system dynamics[17] as described in terms of "stocks" and "flows," while the dynamical evolution of the system is determined by the concepts defined as "forcing" and "feedback." A forcing is an external perturbation to the system which generates a series of enhancing and/or damping feedbacks. Enhancing feedbacks tend to amplify the forcing, damping feedbacks have the opposite effect. The result of enhancing feedbacks may well be to perturb the system to such a point that it crosses a "tipping point"[23] and moves to a different state. This transition can be described as a collapse if the new state corresponds to a condition of lower complexity. In particular, the concept of enhancing feedback may explain the wide variety of attribution of societal collapses to different causes: the initial perturbation which unbalanced the system goes unnoticed when it is masked by the large feedbacks it generates.

In the present study, we use systems dynamics to analyze a well-known biophysical concept that we apply to the concept of civilization: the trophic chain. It assumes that a socio-economic system can be represented as an ecosystem where different trophic levels participate in the degradation, or the dissipation, of the thermodynamic potential associated with the highest potential stock (or lowest trophic level). In the case of civilizations, the lowest trophic level is the one defined as "natural resources," e. g., fossil fuels in the case of the modern civilization. Fossil fuels have a high thermodynamic potential which can be dissipated in the form of the heat produced when fuels combine with atmospheric oxygen. Then, the resulting cascade of trophic levels corresponds to different elements of the economic system: the extractive industry, the manufacturing industry, the bureaucracy, and – lastly – pollution (or waste): the end result of the economic process. This approach is part of the concept of "world modelling" developed first by Jay Forrester[24] and then applied to studies such as the well-known 1972 study titled "The Limits to Growth" [25,26]. A modern example of a world model is the "MEDEAS" model (www.medeas.eu), which aims at describing the whole world system in detail with several hundred parameters. Other models based on the same approach describe the behavior of a complex system by a small number of parameters and may be termed "mind sized"[19]. One of these relatively simple models is the "HANDY" model created by Motesharrei et al. [27]

The model developed here is a simple trophic chain of stocks, in its simplest possible form is a two-stock model, already described in a previous paper[19]. The model is shown in figure 2.

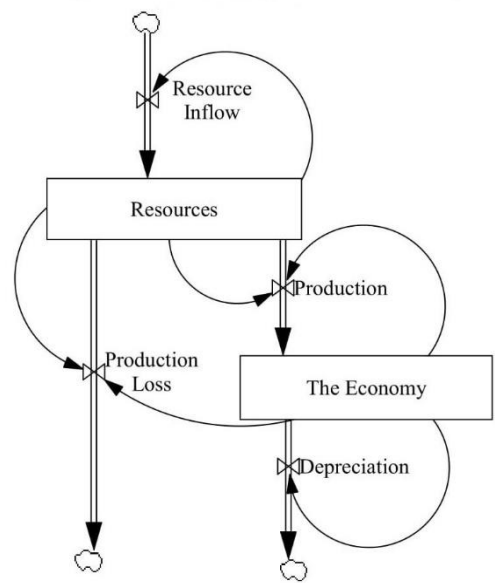

Figure 2: A simple 2-stock System Dynamics model describing the flow and the dissipation of natural resources in a complex society. All the flows in the model are determined by constant multipliers, not shown in the diagram. Note that we represent the flows in the model as going

"down" from higher thermodynamic potentials toward lower thermodynamic potentials. It is a convention described in ref. [19]

The model is represented in a form that emphasizes the unidirectional flows from higher potential stocks to lower potential stocks, going "down" in the arrangement. We can say that the stock termed "The Economy" dissipates the energy potentials contained in the "Resources" stock, turning them into waste heat and non-recyclable materials. For instance, if the resource is crude oil, then the economy transforms it into all the products that can be made using crude oil, fuels, plastics, chemicals, etc. Note also that the model as shown involves no entropic loss for the transformation, that is it assumes a 100% efficient of transformation – such a loss can be easily introduced in the model, but that doesn't change the qualitative results of the model. Note also that the model does not consider the effects of pollution, while other factors, such as human population, social structure, bureaucracy, etc., are all aggregated in the "Economy" stock.

In the model, the flow between stocks is regulated by feedbacks. We assume that each flow is linearly proportional to the size of the stocks connected by it, multiplied by a proportionality constant. It would be also possible to assume that there exist exponents to the stock sizes analogous to the ones called "elasticities" in some macroeconomic models. Here, however, we will use a simple assumption of linear proportionality.

The size of the stocks of the model can be measured in terms of energy: natural resources such as crude oil can be measured in terms of the chemical energy they contain, while other mineral resources can be measured in terms of the "embedded energy" [28] necessary for their extraction, refining, and production. The stocks aggregated in the "economy" stock can also be described in terms of embedded energy – e.g. a piece of machinery can be measured in terms of the energy necessary to create it, including the cost of producing the metals and the other materials needed. The stock sizes can also be measured in monetary units, assumed to be "proxies" of energy units. Here, no quantitative assessment of real-world stocks is attempted, but the fact that they can be all measured in the same units ensures that the transformations described in the model are physically possible.

If the natural resources are supposed to be non-renewable (i.e. the flow into the resource stock is set to zero) this simple two-stock model generates a symmetric, bell-shaped curve for the flow of energy that goes from the upper stock to the lower stock, a flow that we may call "production". This curve corresponds to the "Hubbert curve" which approximately describes the cycle of extraction of mineral resources[29]. If, instead, the natural resource stock is assumed to reform at a rate proportional to the stock size, as it happens for a biological stock, the model is equivalent to the well-known Lotka-Volterra one[30,31] and it generates continuous oscillations in the size of both stocks. Here, we assume that the Resources stock is slowly renewable, so we consider only one cycle of growth and decline.

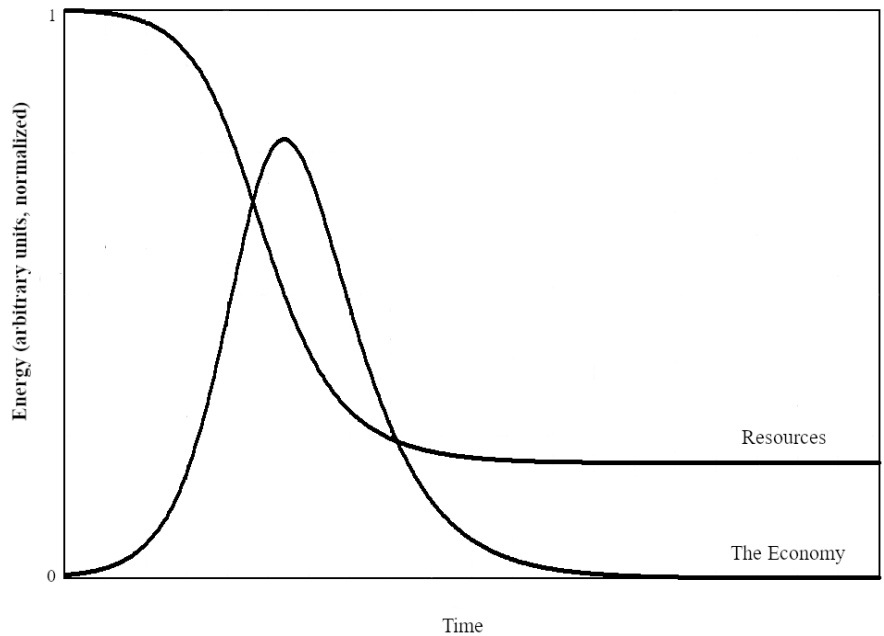

Figure 3: Typical behavior of the stocks in the "2-stock" model for nonrenewable natural resources. The initial values for the stocks are assumed to be 1 for the Resources stock and 0.01 for the Economy stock. The constants determining the flows are 0.1 for the "production" flow and 0.5 for the depreciation flow.

The simple two-stock model is the basis of more complex models where we consider further trophic levels. The next step involves adding a third stock, labelled as "pollution" - this model has been described in a previous paper[18] where it was termed the "Seneca Model" because it produces asymmetric bell-shaped curves where decline is faster than growth, corresponding to an observation put forward long ago by the Roman philosopher Lucius Annaeus Seneca[32].

A further step consists in adding one more level which we term here as "Bureaucracy" supposed to be a stock that aggregates all the non-productive societal structures: the army, the nobility, the court, the priests, etc. The trophic chain, at this point, is the following.

1. Resources
2. The Economy
3. Bureaucracy
4. Pollution

The model is shown in figure 4.

The flow is unidirectional, and it goes in this way: Natural Resources → The Economy → Bureaucracy → Pollution. As in the simpler two-stock model, resources are transformed into economic capital. This capital is partly turned into "bureaucracy" whose damping effect may be taken into account as an indirect effect caused by the drawdown of resources from the economy stock, which is not available to exploit natural resources. Note that this stock is connected to the "production" flow. We may assume that bureaucracy may have an enhancing effect on production. In the real world, this effect could play out, as an example, by providing the extractive industry with a legal framework that allows them to exploit the resource they control without the need of defending them from competitors. This effect can be neglected in a simplified form of the model.

All stocks generate pollution. the necessary result of all the ongoing operation of society. Pollution may take the form of gases emitted by the combustion of fossil fuels, heavy metals dispersed in the environment, as well as the destruction of the fertile soil and the general disturbance of the ecosystem. The pollution stock may be assumed to abate slowly as the polluting substances are re-absorbed by the environment. All stocks in the model are affected by feedbacks which make flows proportional to the size of the stocks connected by it. Note that Bureaucracy affects production in terms of a multiplying factor assumed to be (1+Bureaucracy). This assumption accounts for the fact that Bureaucracy is a less important factor than economic capital in facilitating production.

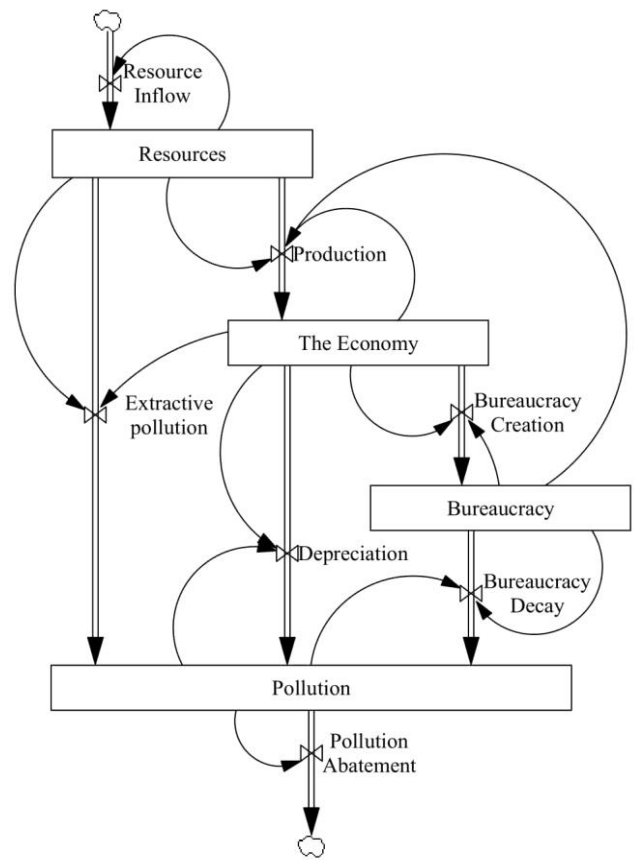

Figure 4: The four-stock SD model used in this paper. All the flows are regulated by constants, not shown in the diagram.

If the resources are assumed to be non-renewable, all the stocks of the system go to zero one after the other. However, the 4-stock model produces a more abrupt collapse than the 2-stock model. This behaviour often takes the form of the "Seneca Collapse"[32]. It is the result of a stock being subjected to a feedback-dominated drawdown by another stock, while at the same time being unable to maintain a replenishing flow from a depleted stock (this can be termed the "candle burning at both ends" effect).

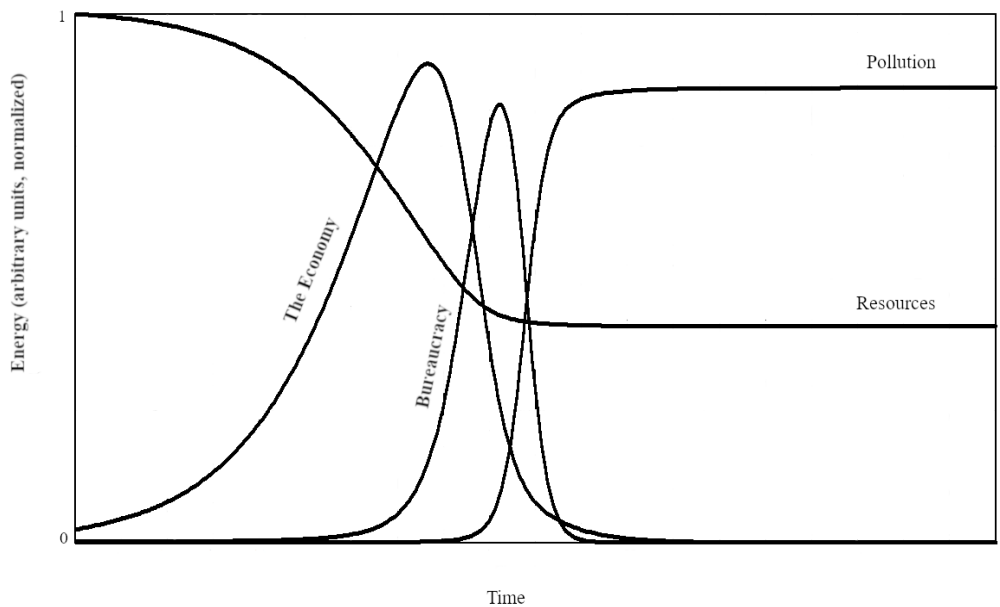

Figure 5: Typical behaviour of the 4-stock model. In this figure, all the four stocks are shown. The initial values are 1 unit for "Resources", 0.1 units for "The Economy," 0.01 units for "Bureaucracy," and 0.001 units for "Pollution". The constants are 0.38 for the "Production" flow, 0.15 for the "Depreciation" flow, 0.3 for the "Bureaucracy Creation Flow" and 0.5 for the "Bureaucracy Decay" flow. All the other flows constants are assumed to be zero – in other words, the resources are assumed to be non-renewable and pollution to be persistent.

A robust feature of the model is the how the decline in the stock of natural resources is related to a cycle of growth and decline of the other stocks. In most cases, the decline turns out to be faster than growth, a characteristic defined here as "collapse." Note also that the behavior of the model can vary depending on the initial assumptions. Depending on the stock parameter, the bureaucratic stock may go to zero before the capital stock does. In this case, we may see the model as describing a civilization that loses some of its centralized control structure (e.g. an imperial court) and moves toward simpler delocalized structures. It could describe how the Middle Ages feudal structure developed after the collapse of the Western Roman Empire. A different behavior may also occur, that is the bureaucracy stock surviving the collapse of the capital stock. In this case, society maintains for a certain time an overgrown central control structure while the productive structures have disappeared. This might be the case of the Eastern Roman Empire which saw the capital city of Constantinople surviving even though most of the territory of the empire had been lost.

The question is now how these models can be related to the qualitative descriptions given by Tainter. We can note first that Tainter describes complexity as[1]:

> Complexity is generally understood to refer to such things as the size of a society, the number and distinctiveness of its parts, the variety of specialized social roles that it incorporates, the number of distinct social personalities present, and the variety of mechanisms for organizing these into a coherent, functioning whole.

From this definition, it seems that we can identify a proxy for the concept of complexity in terms of "The Economy" for the simple 2-stock model and "Bureaucracy" for the case of the four-stock model. In the latter case, the stock is surely composed of those people whom Tainter describes "specialists not directly involved in resource production."

Then, what is that we should see as "benefits of complexity" in the model? The stocks of the system can be replenished only as long as the producing capital can extract energy and materials from the resource stock. Therefore, we can quantify the benefits of complexity using the "production" parameter as a proxy. We can therefore use the model to reproduce Tainter's model of the diminishing returns to complexity by plotting production as a function of the size of the bureaucracy stock. We do this first for the simplified 2-stock model (figure 6) and then for the more complex 4-stock model (figure 7). In both cases, the parameters are those described in the captions for the figures 4 and 5. The results are qualitatively similar.

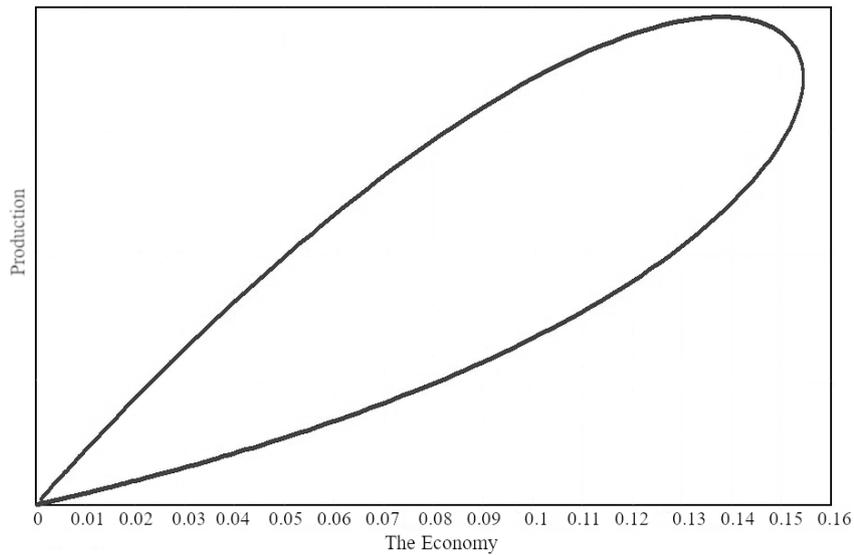

Figure 6: Production vs "The Economy" for the two-stock model. The values of the constant are the same as those described in figure 3.

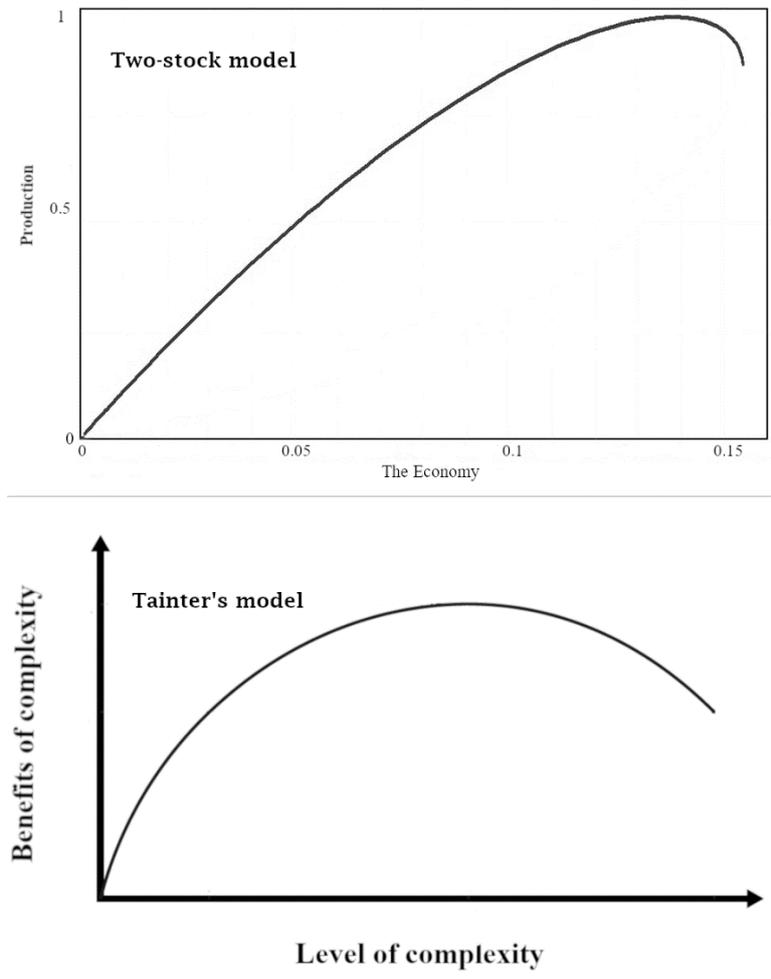

Figure 7, a comparison of the results of the two-stock model reported here and Tainter's model as reported in his 1988 book[1]. The two curves are not identical, but the similarity is evident.

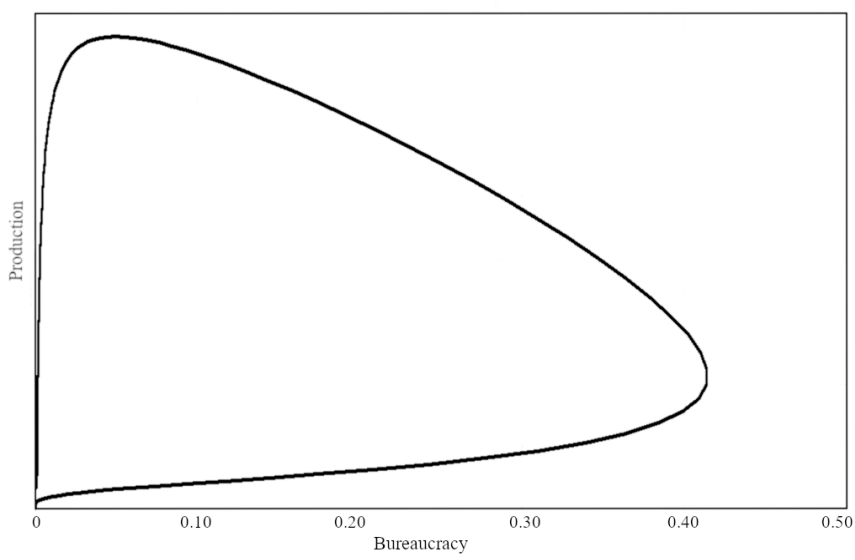

Figure 8: The curve of Production vs. Bureaucracy for the 4-stock model. The values of the constants in the model are the same as those in figure 5. This diagram has a different shape in comparison with the 2-stock model, but the qualitative dependency of the two stocks, bureaucracy and production, remains the same: growth and decline, followed by an inversion in the trend.

The curve for the simpler 2-stock model is qualitatively similar to the one proposed by Tainter, (fig 7) although the two-stock model shows the whole loop (fig. 6) whereas Tainter's illustration shows only the growing phase of complexity. Note the hysteresis of the curve: the bureaucracy stock declines after having reached a maximum value, but its relation to the productivity of the system doesn't go back to the earlier values. In other words, halving the costs of bureaucracy during the decline phase doesn't lead back to the same conditions when the system was growing. The behaviour is qualitatively similar to the 4-stock model although, in this case, Bureaucracy initially favours growth but later on becomes a burden. This phenomenon appears to correspond to the current conditions of modern society. Governments everywhere are cutting their bureaucratic expenses, but the system is not returning to the efficiency of the earlier times.

We may also use a more detailed model to check the behaviour that we observe for our "mind sized" model. A well-known such model is the one called "World3," which was used for the 1972 study "The Limits to Growth." We used the most recent available version of World3 to calculate a Tainter plot for the "business as usual" or "base case" scenario. The world3 model has an "industrial output" that we may consider a proxy for Tainter's "benefits of complexity." It doesn't have a "bureaucracy" or "complexity" stock but, after examining the model, we believe that the parameter defined as "Relative Services Output" can be considered proportional to the size of the bureaucracy in the model and hence as a proxy for Tainter's "complexity" concept. The results are shown in figure 8 and are very similar to those obtained using the simpler models described here: we can see the typical hysteresis of these phenomena.

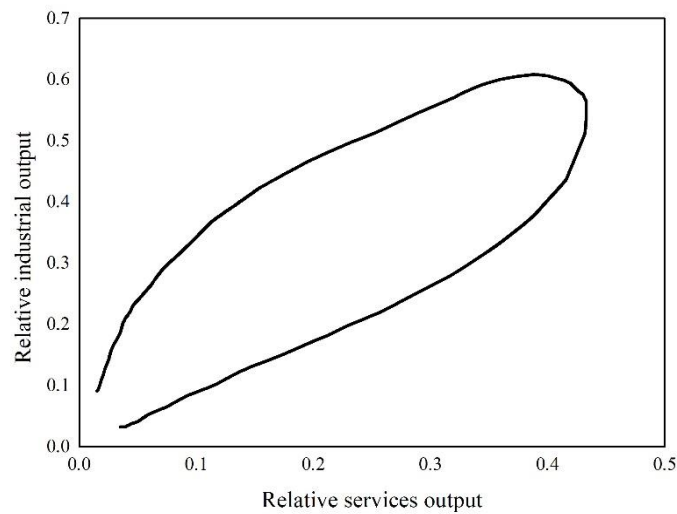

Figure 9: Tainter plot for World 3 where the relative industrial output is reported as a function of the relative services output.

**4. Historical examples.**

In this section, we apply to historical cases the considerations developed in the previous sections. A first example is the Roman empire where the data have been taken from Sverdrup et al[33] who studied the connection between resources depletion and the stability of Roman civilization. In particular, the depletion of silver content in coins that started at the beginning of the 1st century AD continued well into the 3rd century AD and then, decreased. From that moment on, the social structures of empire, such as the army, became so expensive that it was impossible to maintain them at the previous level of organization and size. This phenomenon reflects exactly the concept of diminishing return of complexity: the state becomes larger and more and larger non-producing social structures are needed to maintain it. This is true until the expenses to maintain the bureaucracy and the army does not become too high. In this case, we took the "cash flow" as a measure of the productivity of the Roman Empire and the army size as proportional to the size of the non-productive sector of the Roman economy. Again (figure 10)

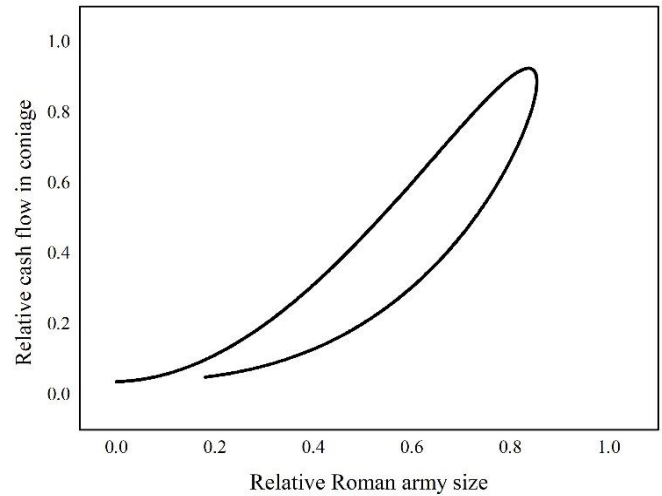

Figure 10: Tainter plot for the Roman empire where the relative cash flow in coinage (state income) is reported as a function of the relative Roman army size. The cash flow is taken as a proxy for Tainter's "benefits of complexity" whereas the army size is taken as a proxy for the complexity of the state system. (data from Sverdrup et al. [34])

A similar behaviour can be observed in the dynamic of exploitation, or better said, in the dynamic of overexploitation of fish. Fish has been, and still is, one of the most precious natural resources that feed the humankind since prehistorical age, placing fishing, and subsequently the fishery sector, as a historical driver in the development of the economy of several countries. However, fisheries are also a good example of the tendency of overexploiting resources, in this case, overfishing[35],[36],[37],[38],[39] . The phenomenon of overfishing has been recently studied by system dynamics in the recent work of Perissi and al.[40] revealing how overshoot is the crucial factor that leads to the destruction of the fisheries. This situation has important consequences in the fisheries economy that can be described exactly by the Tainter's concept of diminishing returns of complexities. The fishing companies tend to invest more and more (fleets, machinery, crews) to stay competitive, a strategy which, incidentally, generates a considerable waste and inefficiency with the fish reaching the market is only about 50%[41] of the total captured. The result is often a dramatic reduction of the population of some species – a phenomenon already observed during the 19th century when the US whale fishery collapse because of overexploitation[42]. More recently, the entire fish population of marine areas has been destroyed or is being destroyed, as it is happening to Japan and Iceland[40].

In the following figures, we report data on fisheries based on the concept of "diminishing returns" proposed by Tainter, taking fish production as a measure of the efficiency of the system and the capital investment as a proxy for the size of the industry (data source Perissi et. Al. [43]). Again, we observe a qualitatively similar behaviour to the others reported before – including the typical hysteresis of the curves. Here, obviously, the fishing industry is not a social system in the same sense the Roman Empire was, but all these systems are evidently similar.

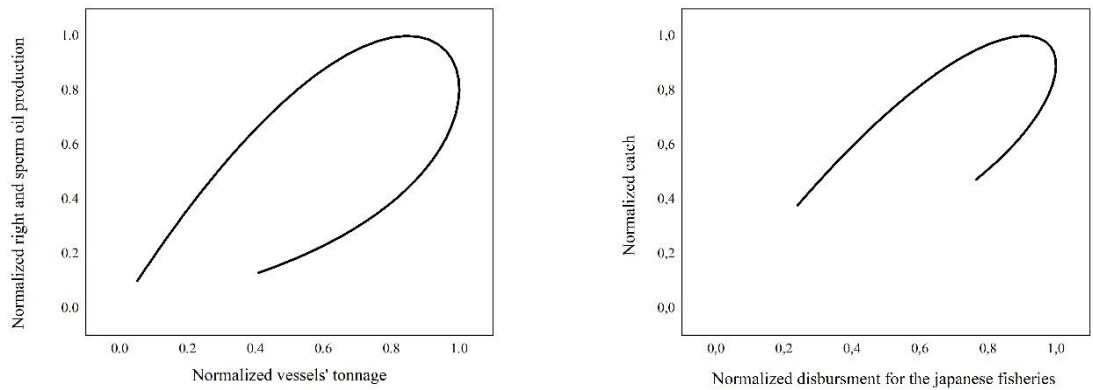

Figure 11: Right. Data for the American Whaling industry during (1820-1880) (data source[44]). Sperm Oil (production) is reported as a function of the Tonnage capacity of fishing boats (complexities) as the effort necessary to maintain the whales' oil production. Left Japanese Fishery Industry (years from 1962 to 2000). Total Catch (production) and the Disbursement of Fishery (complexity) the effort necessary to maintain a competitive landing (data source[43])

## 5. Conclusion

The models studied here are not supposed to describe specific cases of the collapse of human societies, rather, they are thought of as a simplified playground to examine the effect of some parameters on the trajectory of complex systems intended as dissipative structures based on finite or slowly renewable resources. The models are based on a simple concept: that of the trophic chain. If we assume that the natural resources available are non-renewable, as they are in the case of mineral resources (e.g. gold and silver for the Roman Empire and fossil fuels for the modern global empire), the disappearance of the trophic structures exploiting the resources is unavoidable – unless new resources can be found. The same is true for those resources which are slow to renew in comparison to the rate of exploitation. The models tell us how the dissipation of the natural resources goes by the progressive filling and emptying of the stocks at lower thermodynamic potential – every step implying a loss of exploitable potential energy which disappears in the form of pollution, e.g. low temperature heat. This phenomenon generates "bell-shaped" curves for the filling/emptying of the stocks. These curves can also take the "Seneca shape"[32] when the decline is faster than the growth. As this phenomenon goes on, the stocks interact with each other. The time delay in the filling/emptying of the stocks generates a trajectory where stocks move in the phase space along a hysteresis curve. The system continuously evolves in an irreversible manner and it can never return to an earlier condition, unless the resources are assumed to be renewable and, in this case, the system circles around an attractor in phase-space. In other words, simply reducing the size of the "Bureaucracy" stock will not return the system to a condition in which it was during the growing cycle, an observation which seems to correspond to the current situation.

Overall, as long as a society exploits resources in a condition of unbridled feedback, as it happens when it tries to maximize yields, then the overexploitation collapse is unavoidable even though the resources are theoretically renewable. Only an intelligent control able to plan for the future can avoid this destiny. Such a control was not modelled in the present study, but the records of history tell us that it is rarely – if ever – utilized in human societies in history.